\documentclass[aps,prl,reprint]{revtex4-1}
\usepackage[margin=0.77in]{geometry}
\usepackage{amssymb,mathrsfs,physics,amsthm,amssymb,amsfonts,amsmath}

\usepackage{import}

\usepackage{hyperref}
\hypersetup{
    plainpages=false,       
    unicode=false,          
    pdftoolbar=true,        
    pdfmenubar=true,        
    pdffitwindow=false,     
    pdfstartview={FitH},    
    pdfnewwindow=true,      
    colorlinks=true,        
    linkcolor=Blue,         
    citecolor=Blue,        
    filecolor=magenta,      
    urlcolor=blue           
}
\usepackage[usenames, dvipsnames]{xcolor}

\newcommand\id{\leavevmode\hbox{\small1\kern-3.3pt\normalsize1}}

\usepackage{enumitem}

\usepackage[normalem]{ulem}

\theoremstyle{definition}
\newtheorem{theorem}{Theorem}

\usepackage{upgreek}

\usepackage{graphicx}
\graphicspath{ {images/} }

\usepackage{caption}

\usepackage{tabularx}

\usepackage{bbm}

\usepackage{tikz}
\usetikzlibrary{arrows}

\usepackage{CJKutf8}

\usepackage{versions}
\excludeversion{cmt}

\newcommand{\cps}{\mathrel{\ooalign{$\nearrow$\cr \kern-0pt$\nwarrow$}}}

\begin{document}
\title{Reduction of correlations by quantum indefinite causal structure}
\begin{CJK*}{UTF8}{gbsn}
\author{Ding Jia (贾丁)}
\email{ding.jia@uwaterloo.ca}
\affiliation{Department of Applied Mathematics, University of Waterloo, Waterloo, Ontario, N2L 3G1, Canada}
\affiliation{Perimeter Institute for Theoretical Physics, Waterloo, Ontario, N2L 2Y5, Canada}

\begin{abstract}
We show that quantum indefinite causal structure generically reduce two-party correlations. For significant indefiniteness in the causal structure captured by some general conditions, the correlation is shown to be reduced down to zero. The result offers an operational model-independent ultraviolet regularization mechanism. It may be applied to various approaches of quantum gravity that allow quantum indefinite spacetime causal structure.
\end{abstract}

\maketitle
\end{CJK*}

\textbf{Introduction.} There are several reasons for the general presence of indefinite causal structure in nature: 1) The equivalence principle implies that all forms of matter gravitate \cite{einstein2003meaning}. Quantum matter with quantum uncertainty is expected to gravitate with uncertainty, leading to indefinite causal structure. 2) Thought experiments  (e.g. \cite{salecker1958quantum, jack1994limit, amelino1994limits, padmanabhan1987limitations, garay1995quantum, hossenfelder2013minimal}) of length and duration measurements imply fluctuations in signal arrival times. Events defined by signal arrivals can have indefinite causal structure with events defined by clock time readings. 3) In quantum theory all dynamical variables exhibit uncertainties. According to general relativity, causal structure is dynamical, so is expected to exhibit indefiniteness \cite{hardy2005probability, hardy2007towards}. 4) Information processing protocols using indefinite causality had been conceived (e.g. \cite{hardy2009quantum, chiribella2013quantum, chiribella2012perfect, araujo2014computational, feix2015quantum, guerin2016exponential}) and reportedly realized in experiments \cite{maclean2017quantum, procopio2015experimental, rubino2017experimental}. Artificial setups can create indefinite causal structure in nature.

In this letter, we connect indefinite causal structure to the old problem of ultraviolet (UV) regularization, which is of crucial importance to quantum field theory \cite{weinberg1995quantum}, field entanglement and emergent spacetime programs \cite{sorkin1983entropy, bombelli1986quantum, susskind1994black, bianchi2014architecture, jacobson1995thermodynamics, chirco2010nonequilibrium, padmanabhan2010thermodynamical, van2010building}, and the Hawking and Unruh effects \cite{jacobson1991black, helfer2003black, nicolini2011minimal}. There are some common limitations in some previous studies of natural UV regularizations: 1) The use of specific modified propagators. 2) The reference to a classical background spacetime. 3) The assumption of particular microscopic structures of spacetime. Regarding 1), it is possible to study specific modified propagators and do fairly concrete calculations to explore consequences, but different forms of the modified propagator lead to different predictions, even for the crucial question of whether the Unruh and Hawking effects exist \cite{agullo2008two, agullo2009insensitivity, helfer2010comment, nicolini2011minimal, hossain2015violation, hossain2016there}. Since quantum gravity induced modifications may have a dynamical origin \cite{garay1995quantum} and the modifications may vary from region to region, is it possible to approach quantum gravitational modifications in a more general way? Regarding 2), quantum gravity induced regularization should ideally be studied over quantum spacetime. Effective descriptions over a classical spacetime may introduces new subtleties such as preferred reference frames (e.g., \cite{ford1995gravitons}). It is hard to judge whether these are genuine effects of nature or artificial defects due to the compromise in modelling quantum spacetime classically. Is it possible to simply avoid referring to classical spacetime? Regarding 3), popular approaches to quantum gravity such as loop quantum gravity and string theory already provide robust UV regularization mechanisms \cite{oriti2009approaches}, but these are based on particular (and tentative) assumptions and models about the microscopic structure of spacetime. Is it possible to study UV regularization model-independently?

We answer these questions in the affirmative and propose an indefinite causal structure induced natural UV regularization mechanism. The regularization applies to families of correlations rather than specific ones. The characterization of the correlations and the proof of the results are based on an operational approach that is free from referring to a classical background spacetime and from committing to particular assumptions and models about the microscopic structure of spacetime. 

Technically, we show that generically indefinite causal structure reduces two-party correlations. We identify sufficient conditions that reduce the correlations quantified by coherent information (negative conditional entropy) based measures down to zero. The conditions capture major known correlations with significant indefiniteness in the causal structure. Because the coherent information based measures are continuous, correlations close to the above ones but with weaker indefiniteness in the causal structure have correlations close to zero. In this sense, indefinite causal structure reduces correlations generically. To understand the result as a UV regularization, consider two arbitrary operationally meaningful detectors that probe the quantum field correlations of two regions. As quantum devices displaying finitely many possible classical outcomes, the detectors will be described by some finite dimensional quantum instruments \cite{davies1970operational}. Under some general physicality conditions, their correlations will be described by the process matrices \cite{oreshkov2012quantum}. Without quantum gravitational effect and indefinite causal structure, i.e., without a natural UV regularization mechanism, as the detectors approach each other their correlations are allowed to become relatively large since the field correlations become very strong as the detectors get close. If quantum gravitational fluctuations of causal structure is incorporated at short length scales such that the indefiniteness in causal structure becomes significant enough to fall under the sufficient conditions, then the results of this letter show that the correlations must drop down to zero.

\textbf{Process matrices.}\label{sec:pm} The process matrix framework introduced by Oreshkov, Costa and Brukner \cite{oreshkov2012quantum} offers a general way to describe correlations with indefinite causal structure. We give a very brief review below, referring the readers to the original article for details. The basic idea is to take local operations to be described by ordinary quantum theory with definite causal structure, and introduce indefinite causal structure in the global correlations of the local operations.

The local parties where operations are applied are denoted $A, B, \cdots$. Each party $X$ is associated with an input system with Hilbert space $\mathcal{H}^{x_1}$ where information propagates in, and an output system with Hilbert space $\mathcal{H}^{x_2}$ where information propagates out. The parties can share correlations on these systems. For example, a state $\rho$ shared at $A$'s input and $B$'s input is denoted $\rho^{a_1b_1}$, and a channel $N$ from $A$'s output to $B$'s input is denoted $N_{a_2}^{b_1}$ (Following \cite{hardy2011reformulating, hardy2012operator} we used superscripts and subscripts to distinguish input and output systems.). A process $W$ shared by $A$ and $B$ is denoted $W^{a_1b_1}_{a_2b_2}$. It generalizes states and channels to incorporate the correlations among all the input and output systems.


All these objects with inputs and outputs can be represented as operators on Hilbert spaces using the well-known Choi isomorphism \cite{choi1975completely}. The Choi operator of a object is obtained by inputting a maximally entangled state in a canonical basis on each input (which yields a state described by a density operator). Expressed as a matrix in a canonical basis, the Choi operator of a process $W$ becomes a ``process matrix'' denoted by the same symbol. The matrix obeys
\begin{align}
W&\ge 0,
\\
\Tr[W]&=1,
\\
L_V(W)&=W.
\end{align}
These follow from some very general physicality conditions on the correlated outcome probabilities (the first from the non-negativity of probabilities, and the last two from the normalization of probabilities\footnote{Some other works use for the second condition $\Tr[W]=d_O$, where $d_O$ is the dimension of all the output systems taken together. We use the alternate convention of \cite{jia2017quantum, jia2017generalizing} and absorb the factor $d_O$ into the composition rule. This is merely a choice of convention and does not make a difference for the physics.}). In the last line $L_V$ is the projector given in \cite{araujo2015witnessing}. For this paper the details of this projector is irrelevant, except that it implies
\begin{align}\label{eq:norm}
W^{a_1}_{a_2b_2}=W^{a_1}_{b_2}\otimes W_{a_2}.
\end{align}

Objects with sub- and superscripts can compose when the output of one object is fed into the input of another. Such a composition is shown with repeated sub- and superscripts, e.g., sequentially composing the channels $M$ and $N$ yields a new channel $L$: $M_a^b N_b^c=L_a^c$. We use the convention that a discarded system has its label eliminated, e.g., $\rho^a=\Tr_b\rho^{ab}$. In addition, when no ambiguity arises we sometimes omit the labels or refer to objects by the relevant parties, e.g., $W^{a_1b_1}_{a_2b_2}$ is sometimes referred to as $W$ or $W^{AB}$.

The following two examples had been conceived to describe indefinite causal structure that arise in quantum spacetime fluctuations. The first example \cite{maclean2017quantum, jia2017quantum} is a mixture of two causal relations $A\rightarrow B$ ($A$ causally precedes $B$) and $A-B$ ($A$ causally disconnected with $B$). Assume $\dim a_1=\dim a_2=\dim b_1=\dim b_2$. The ``partial swap'' channel $P(p)_{a_1b_1}^{a_2b_2}$ ($0\le p\le 1$) is the channel corresponding to the partial swap unitary:
\begin{align}
\sqrt{1-p}~\id+\sqrt{p}~ i ~ U_{SW}.
\end{align}
The identity $\id$ part sends $a_1$ to $a_2$ and $b_1$ to $b_2$, whereas the swap unitary $U_{SW}$ part sends $a_1$ to $b_2$ and $b_1$ to $a_2$.
\begin{align}
W^{a_1b_1}_{a_2}:=&\Tr_{e_1} P(p)_{a_1'a_2'}^{b_1 e_1}\rho^{a_1 a_1'} N^{a_2'}_{a_2},\label{eq:2crgp}
\\
W^{AB}=&W^{a_1b_1}_{a_2}\otimes \pi_{b_2},\label{eq:cspm}
\end{align}
where $a_1'$ and $a_2'$ are copies of $a_1$ and $a_2$, and $\pi$ is the maximally mixed density operator. The process $W$ puts $A$ and $B$ into a ``coherent superposition'' of sharing an acausal state $\rho$ and a causal channel $N$.

The second example \cite{feix2017quantum, jia2017quantum, jia2018analogue, jia2018quantifying} is a mixture of all the three causal relations $A\rightarrow B$, $A\leftarrow B$ and $A-B$. 
\begin{align}
\ket{w(\alpha)}^{GABE}=&\alpha_1\ket{1}^g\ket{\Psi(\alpha)}^{a_1e_2e_3}\ket{I}^{a_2 b_1}\ket{I}^{b_2 e_1}\nonumber
\\
+&\alpha_2\ket{2}^g\ket{\Psi(\alpha)}^{e_1b_1e_3}\ket{I}^{b_2 a_1}\ket{I}^{a_2 e_2}\nonumber
\\
+&\alpha_3\ket{3}^g\ket{\Psi(\alpha)}^{a_1b_1e_3}\ket{I}^{a_2 e_1}\ket{I}^{b_2 e_2},\label{eq:3crps}
\\
W^{GABE}(\alpha)=&\ketbra{w(\alpha)}^{GABE},
\\
W^{AB}(\alpha)=&\Tr_{GE} W^{GABE}(\alpha).\label{eq:pm3cr}
\end{align}
The party $E$ is the ``environment'' that collects information not collected by the other parties. $\ket{\Psi(\alpha)}$ is a tripartite state depending on the parameter $\alpha$, and $\ket{I}^{xy}$ is vector of the Choi operator $\ketbra{I}^{xy}$ for the identity channel from $x$ to $y$. The party $G$'s system $g$ has basis vectors $\ket{0}$, $\ket{1}$ and $\ket{2}$ from which the causal relations $A\rightarrow B$, $A\leftarrow B$ and $A-B$ can be read respectively. For example, for the $\ket{1}$ term $A$ causally precedes $B$ through sharing the channel $\ket{I}^{a_2b_1}$. $g$ can be thought of as containing quantum gravitational degrees of freedom that induce different causal relation for $A$ and $B$. $\ket{w(\alpha)}$ puts the relations into a ``superpostion''. The probability amplitudes form a complex $3$-vector $\alpha=(\alpha_1,\alpha_2,\alpha_3)$ with $\norm{\alpha}_2=1$. 

\textbf{Measures of correlation and some lemmas.} In the following $S^a$ denotes the von Neumann entropy of a density operator on system $a$. A widely used measure of correlation is the mutual information. For a state $\rho^{ab}$, it is defined as $I^{a:b}(\rho)=S^a+S^b-S^{ab}$, where $S^a$ ($S^b$) is the von Neumann entropy of $\rho^a$ ($\rho^b$). The mutual information measures both quantum and classical correlations. To measure only quantum correlations, the coherent information can be used. It is defined as $I^{a}(\rho^{ab})=S^a-S^{ab}$ ($I^{b}(\rho^{ab})=S^b-S^{ab}$) when the target system is $a$ ($b$). The coherent information is exactly the negative of the conditional entropy, and differs from the mutual information only by $S^b$ ($S^a$). It can be positive only if the state is entangled, and attains the maximum value for maximally entangled states. For pure states the coherent information coincides with the entanglement entropy. Although for general states the coherent information is not an entanglement measure since local operations and classical communications (LOCC) may increase it, the LOCC-optimized coherent information $I^{a}_{\text{LOCC}}(\rho):=\sup_{L\in\text{LOCC}}I^{a'}(L(\rho))$ is. Here the optimization is over the set $\text{LOCC}$ of the allowed LOCC operations. Depending on the context, different sets of LOCC operations (e.g., two-way classical communication, one-way classical communication, no classical communication) will be allowed. $a'$ with a prime shows up because $L$ (e.g., by changing system dimensions) may map to a Hilbert space different from the original $a$.

These ``coherent information based measures'' for states can be generalized to process matrices \cite{jia2017generalizing}. In this letter we use the coherent information $I^B(W^{AB}):=S^B-S^{AB}$ and the LO-optimized coherent information $I^{B}_{\text{LO}}(W^{AB}):=\sup_{L\in\text{LO}} I^{B}(L(W^{AB}))$. $S^{B}$ and $S^{AB}$ are the von Neumann entropies of the process matrix reduced to the party $B$ and of the whole process matrix. The optimization is over the set $\text{LO}$ of local operations without classical communication, because in the present context we are interested in indefinite causal structure, and allowing classical communication would make the causal structure become trivially causal connected. No prime needs to be introduced on $B$ because it refers to whatever system the party $B$ obtains after the optimization. The process matrices are treated as density operators for the evaluation. The state represented by the density operator can be operationally obtained from the process by inputting a maximally entangled state to each input system. Hence the above measures are interpreted as the the coherent information of the corresponding states obtained from the processes. The latter optimized measure is an entanglement measure in the generalized sense \cite{jia2017generalizing}.

We will make use of the following standard results from quantum information theory \cite{wilde2017quantum}.
\begin{theorem}[Subadditivity]\label{th:suba}
For a density operator $\rho^{ab}$, 
\begin{align}\label{eq:sa}
S^{a}+S^{b}\ge S^{ab}.
\end{align}
\end{theorem}

\begin{theorem}[Strong subadditivity]\label{th:ssuba}
For a density operator $\rho^{abc}$, 
\begin{align}\label{eq:ssa}
S^{ab}+S^{bc}\ge S^{abc}+S^{b}.
\end{align}
\end{theorem}

\begin{theorem}\label{th:cienv}
For a bipartite density operator $\rho^{ab}$ purified by $\rho^{abe}$, 
\begin{align}
I^b(\rho^{ab})=-I^e(\rho^{ae}).
\end{align}
\end{theorem}

\begin{theorem}[Data processing inequality for coherent information]\label{th:dpici}
For any bipartite state $\rho^{ab}$ and any channel $N_b^c$,
\begin{align}
I^b(\rho^{ab})\ge  I^c(N_b^c \rho^{ab}).    
\end{align}
\end{theorem}

\begin{theorem}[Alicki-Fannes-Winter inequality]\label{th:afwi}
For density operators $\rho^{ab}$ and $\sigma^{ab}$, if $\frac{1}{2}\norm{\rho^{ab}-\sigma^{ab}}_1\le \epsilon$ for $\epsilon\in [0,1]$. Then
\begin{align}
&\abs{I^{b}(\rho)-I^{b}(\sigma)}\le 2\epsilon \log \abs{a}+(1+\epsilon)h_2(\epsilon/(1+\epsilon)),
\end{align}
where $\abs{a}$ is the dimension of system $a$ and $h_2(p):=-p\log p-(1-p)\log(1-p)$ is the binary entropy.
\end{theorem}

\textbf{Main results.}\label{sec:mr} 
$A$ and $B$ can have three possible definite causal relations $A\rightarrow B$, $A-B$, and $A\leftarrow B$. There are four possible ways to (quantum coherently or classically) mix these relations: three ways to mix two relations, and one way to mix three relations. Among these, the mixture of $A\rightarrow B$ with $A\leftarrow B$ is not expected to take place through naturally occurring quantum gravitational fluctuations, because it leaves out the intermediate case $A-B$. In addition, the mixture of $A\rightarrow B$ with $A-B$ and that of $A\leftarrow B$ with $A-B$ are of the same type. Therefore we restrict attention to two cases, the mixture of  $A\rightarrow B$ with $A-B$ and that of all three relations.

The following theorems identify sufficient conditions to reduce coherent information based measures down to zero. By the continuity of the coherent information (Theorem \ref{th:afwi}), process matrices close to these characterized by the conditions have measures close to zero. The intuition behind the reduction of correlation is that indefinite causal structure induces leakage of correlation into the environment (This is the intuitive meaning of Theorem \ref{th:dpici} and \ref{th:ssuba} used in the proofs.). 

For the indefinite causal structure of $A-B$ with $A\rightarrow B$, we have the following theorem.
\begin{theorem}\label{th:2crci0}
Let $W^{AB}$ be a process matrix of the form
\begin{align}
W^{AB}=W^{a_1 b_1}_{a_2}\otimes W_{b_2}.\label{eq:2crcond1}
\end{align}
Suppose there is a $\ketbra{w}^{ABE}=W^{a_1 b_1 e_1}_{a_2}\otimes W_{b_2}^{e_2}$ that purifies $W^{AB}$ as a density operator, so that for a subsystem $e_0$ of $e_1$,
\begin{align}
W^{a_1 b_1}_{a_2}=W^{a_1 e_0}_{a_2}.\label{eq:2crcond2}
\end{align}
Then $I^{B}_{LO}(W^{AB})=0$.
\end{theorem}

In example (\ref{eq:cspm}) the most significant indefinite causal structure comes with $p=1/2$, which has equal probability amplitudes $A\rightarrow B$ and $A-B$. By taking $e_0$ to be $e_1$ itself, the above conditions are met. In example (\ref{eq:pm3cr}) purified by (\ref{eq:3crps}), the most significant indefinite causal structure for mixing $A-B$ with $A\rightarrow B$ comes with $\alpha_1=\alpha_3=1/\sqrt{2}$ and $\alpha_2=0$. The above conditions are met under the relabelling ($e_1=ge_0e_3$):
\begin{align}
\ket{w(\alpha)}^{GABE}=&\alpha_1\ket{1}^{g}\ket{\Psi(\alpha)}^{a_1e_0e_3}\ket{I}^{a_2 b_1}\ket{I}^{b_2 e_2}\nonumber
\\
+&\alpha_3\ket{3}^g\ket{\Psi(\alpha)}^{a_1b_1e_3}\ket{I}^{a_2 e_0}\ket{I}^{b_2 e_2}.
\end{align}
In general, (\ref{eq:2crcond1}) says that there is no $A$ to $B$ signal term \cite{oreshkov2012quantum}, which holds when $A\rightarrow B$ mixes with $A-B$. Condition (\ref{eq:2crcond2}) reflects significant indefinite causal structure. $A$ has equal chance to be causally prior to $B$ (so $e_0$ is causally disconnected with $A$) and causally disconnected with $B$ (so $e_0$ is causally prior to $A$), such that $B$ and the environment $e_0$ share correlations with $A$ in the same way.
\begin{proof}
The local operations map the original process to a new one, which is a density operator. The state described by the same density operator can be obtained from the new process by sending maximally entangled states to every input. Hence to evaluate the coherent information of the density operator we can without loss of generality assume that the LO already maps to the state. Denote by $M^{a_2a'}_{a_1}$ and $N^{b_2b'}_{b_1}$ some arbitrary such local operations conducted by $A$ and $B$. Then
\begin{align}
\rho^{a'b'}:=&M^{a_2a'}_{a_1}N^{b_2b'}_{b_1}W_{a_2b_2}^{a_1b_1}
\\
=&M^{a_2a'}_{a_1}N^{b_2b'}_{b_1} W_{a_2}^{a_1b_1}\otimes \Tr_{e_2} W_{b_2}^{e_2}
\\
=&(\Tr_{e_2} W_{b_2}^{e_2} N^{b_2b'}_{b_1}) (M^{a_2a'}_{a_1} W_{a_2}^{a_1b_1}).
\end{align}
This is a channel $T_{b_1}^{b'}$ in the first bracket applied to a state $\omega^{a'b_1}$ in the second bracket. By the data-processing inequality (Theorem \ref{th:dpici}),
\begin{align}
I^{b'}(\rho^{a'b'}):=&I^{b'}(T_{b_1}^{b'}\omega^{a'b_1})
\le I^{b_1}(\omega^{a'b_1}). \label{eq:2crst1}
\end{align}

Let $M^{a_2a'e_3}_{a_1}$ be an isometric extension of $M^{a_2a'}_{a_1}$. Since $W_{a_2}^{a_1b_1e_1}$ is pure, $\omega^{a'b_1e_1e_3}=M^{a_2a'e_3}_{a_1} W_{a_2}^{a_1b_1e_1}$ is a purification of $\omega^{a'b_1}=M^{a_2a'}_{a_1} W_{a_2}^{a_1b_1}$. Hence
\begin{align}
I^{b_1}(\omega^{a'b_1})=&-I^{e_1e_3}(\omega^{a'e_1e_3})\le-I^{e_0}(\omega^{a'e_0}).
\end{align}
The equality uses Theorem \ref{th:cienv}, and the inequality uses Theorem \ref{th:dpici} for which the post-processing channel traces out the complement of $e_0$ in $e_1e_3$. Now by (\ref{eq:2crcond2}),
\begin{align}
I^{b_1}(\omega^{a'b_1})=I^{e_0}(\omega^{a'e_0}).
\end{align}
Therefore $I^{b_1}(\omega^{a'b_1})=I^{e_0}(\omega^{a'e_0})\le -I^{e_0}(\omega^{a'e_0})$. It must be that $I^{b_1}(\omega^{a'b_1})=0$. By (\ref{eq:2crst1}), this implies that $I^{b'}(\rho^{a'b'})\le 0$. Since $M$ and $N$ are arbitrary, the optimized value $I^{B}_{LO}(W^{AB})\le 0$. We also know that the value cannot be negative (with local operations that discard everything and jointly prepare a product of pure states, the value is zero), so it is $0$.
\end{proof}

Next we consider indefinite causal structure of all three relations.
\begin{theorem}\label{th:3crer}
Let $W^{AB}$ be a process matrix, whose density operator is purified by $W^{a_1 b_1e}_{a_2 b_2}$. Suppose $W^{a_1 b_1}_{a_2 b_2}$ is symmetric in $A$ and $B$. Suppose further that $e$ has a subsystem $e_1$ such that $S^{b_1}\le S^{e_1}$ and $S^{a_1b_2}\ge S^{b_1e_1}$. Then $I^{B}(W^{AB})\le 0$.
\end{theorem}
An example of a process matrix satisfying the conditions is (\ref{eq:pm3cr}) with $\alpha_1=\alpha_2=\alpha_3=1/\sqrt{3}$ and $\ket{\Psi(\alpha)}^{xyz}=\ket{\Upphi_+}^{xy}$ (a maximally entangled state). The probability amplitudes of the three causal relations are equal, so the causal structure is significantly indefinite. In general, $W^{AB}$ is symmetric in  $A$ and $B$ if indefinite causal structure washes out any asymmetry that comes from an initial definite causal structure. 
The other condition can be heuristically interpreted as saying that an environmental subsystem $e_1$ is correlated with $b_1$ no less strongly than $a_1$ is correlated with $b_2$ ($S^{a_1b_2}\ge S^{b_1e_1}$), while obeying the technical condition $S^{e_1}\ge S^{b_1}$. 

\begin{proof}
Writing $a$ ($b$) for $a_1a_2$ ($b_1b_2$), we have
\begin{align}
I^{B}(W^{AB})=&S^b-S^{ab}
\\
=&S^b-(S^{a b_2}+S^{ab}-S^{ab_2})
\\
=&S^b-(S^{a b_2}+S^{e}-S^{b_1e})\label{eq:uwp}
\\
\{\text{(\ref{eq:norm}})\}=&S^b-(S^{a_1 b_2}+S^{a_2}+S^{e}-S^{b_1e})
\\
\{S^{a_1b_2}\ge S^{b_1e_1}\}\le&S^b-(S^{b_1 e_1}+S^{a_2}+S^{e}-S^{b_1e})
\\
\{(\ref{eq:sa})\}\le &(S^{b_1}+S^{b_2})-(S^{b_1 e_1}+S^{a_2}+S^{e}\nonumber\\&-S^{b_1e})\label{eq:usa}
\\
\{S^{a_2}=S^{b_2}\}=&S^{b_1}-(S^{b_1 e_1}+S^{e}-S^{b_1e})
\\
\{S^{b_1}\le S^{e_1}\}\le&S^{e_1}-(S^{b_1 e_1}+S^{e}-S^{b_1e})
\\
=&S^{e_1}-S^{b_1 e_1}-S^{e_1\bar{e}}+S^{b_1e_1\bar{e}}
\\
\{(\ref{eq:ssa})\}\le & 0.
\end{align}
The justifications for the non-trivial steps are written in the curly braces. Line (\ref{eq:uwp}) holds because $W^{abe}$ is pure. $S^{a_2}=S^{b_2}$ because $W$ is symmetric in $A$ and $B$. In the second to last line $\bar{e}$ is introduced as a complementary system such that $e=e_1\bar{e}$.
\end{proof}
Whether the LO-optimization can increase the measure to a some positive number is an open question that deserves further attention. Finally, the mutual information $I^{a:b}$ is related with the coherent information by $I^{a:b}:=S^a+S^b-S^{ab}=I^a+S^b$. For each coherent information based measure there is corresponding mutual information based measure (taking into account classical correlation) defined by substituting $I^{a:b}$ for $I^a$. The reduction of the coherent information based measures implies the reduction of the mutual information based measures.

\textbf{Discussion.} In this letter we identified some sufficient conditions for indefinite causal structure to reduce two-party correlations to zero. Since the measures of correlation are continuous, the reduction of correlation by indefinite causal structure is a generic effect. Applied to quantum fluctuations of spacetime causal structure, this effect offers a UV regularization mechanism.

This work can be developed further. Is there a more general condition that characterize correlations with significantly indefinite causal structure than the sufficient conditions we found? Can local operations increase the coherent information in Theorem \ref{th:3crer}? Is there a theory of field propagators with indefinite causal structure that allows more concrete calculations of field detectors?

\section*{Acknowledgement}
I am very grateful to Lucien Hardy and Achim Kempf for guidance and support as supervisors, and to Jason Pye for valuable discussions.

Research at Perimeter Institute is supported by the Government of Canada through the Department of Innovation, Science and Economic Development Canada and by the Province of Ontario through the Ministry of Research, Innovation and Science.  This work is supported by a grant from the John Templeton Foundation. The opinions expressed in this work are those of the author's and do not necessarily reflect the views of the John Templeton Foundation.

\bibliographystyle{apsrev} 
\bibliography{bibliography} 

\end{document}